\documentclass[12pt,preprint]{emulateapj}
\usepackage{natbib}
\shorttitle{On The Origin Of The Mass-Metallicity Relation For GRB Host Galaxies}
\shortauthors{D. Kocevski et al.}

\begin{document}

\title{On The Origin Of The Mass-Metallicity Relation For GRB Host Galaxies}

\author{Daniel Kocevski \altaffilmark{1}, Andrew A. West \altaffilmark{2} }

\altaffiltext{1}{Kavli Institute for Particle Astrophysics and Cosmology, Stanford University, 2575 Sand Hill Road M/S 29, Menlo Park, Ca 94025 }
\altaffiltext{2}{Department of Astronomy, Boston University, CAS 422A, 725 Commonwealth Ave, Boston, MA 02215}


\begin{abstract}

We investigate the nature of the mass-metallicity ($M$-$Z$) relation for long gamma-ray burst (LGRB) host galaxies.  Recent studies suggest that the $M$-$Z$ relation for local LGRB host galaxies may be systematically offset towards lower metallicities relative to the $M$-$Z$ relation defined by the general star forming galaxy (SDSS) population.  The nature of this offset is consistent with suggestions that low metallicity environments may be required to produce high mass progenitors, although the detection of several GRBs in high-mass, high-metallicity galaxies challenges the notion of a strict metallicity cut-off for host galaxies that are capable of producing GRBs.  We show that the nature of this reported offset may be explained by a recently proposed anti-correlation between the star formation rate (SFR) and the metallicity of star forming galaxies. If low metallicity galaxies produce more stars than their equally massive, high-metallicity counterparts, then transient events that closely trace the SFR in a galaxy would be more likely to be found in these low metallicity, low mass galaxies.  Therefore, the offset between the GRB and SDSS defined $M$-$Z$ relations may be the result of the different methods used to select their respective galaxy populations, with GRBs being biased towards low metallicity, high SFR, galaxies.  We predict that such an offset should not be expected of transient events that do not closely follow the star formation history of their host galaxies, such as short duration GRBs and SN Ia, but should be evident in core collapse SNe found through upcoming untargeted surveys.

\end{abstract}


\section{Introduction}

Investigations of the environments in which gamma-ray bursts (GRBs) occur has long been an important path to understanding the nature of their progenitors, as different progenitor models have traditionally predicted distinct GRB host galaxy populations.  The observations of long (LGRB) host galaxies, made possible through X-ray localizations of GRB afterglows, show that they tend to be bluer, fainter, and later type than $M_{\star}$ galaxies at similar redshifts \citep{Fruchter99, Chary02, Bloom02, LeFloch03, Tanvir04, Fruchter06, Ceron06}.  They also tend to have higher specific star formation than typical star-forming galaxies \citep{Chary02, Berger03, Christensen04} and their redshift distribution tends to broadly track the overall cosmic star formation rate (SFR) of the universe \citep{Bloom03, Firmani04, Natarajan05, Jakobsson06,  Kocevski06, Guetta07}.  These results have bolstered the connection between GRBs and the death of massive stars, which is now well-established at low redshift ($z < 0.3$) through the association of GRBs with broad lined SN lc events \citep[for a review, see][]{Woosley06}.

Spectroscopic observations of host galaxies have also shown that they tend to be metal poor \citep{Prochaska04, Sollerman05, Fruchter06, Modjaz06, Stanek07, Thoene07, Wiersema07, Margutti07}.  A detailed comparison between the metallicity at the sites of broad-lined SN Ic that have been associated with GRBs and the site of SN Ic with no detected gamma-ray emission found that the chemical abundance of SN-GRB hosts were systematically lower than the hosts of SN without GRBs  \citet{Modjaz08}. This metallicity difference raised the possibility of a upper limit to the metallicity of a galaxy that can produce a GRB. 

Recently, \citet{Han10} and \citet{Levesque10} compared the mass-metallicity ($M$-$Z$) relation for long GRB host galaxies to samples from the Sloan Digital Sky Survey (SDSS) representative of the general star-forming galaxy population \citep{Tremonti04}.  Using a small sample of 5 host galaxies, \citet{Han10} found that the metallicities of the host galaxies tended to fall below the low redshift $M$-$Z$ relation defined by SDSS catalog.  Likewise, \citet{Levesque10} compared a much broader sample of LGRB host galaxies and found a similar offset, with LGRB host galaxies exhibiting lower metallicities compared to SDSS galaxies of similar masses.

The relative nature of this metallicity offset for a given host mass, along with a small but growing number of high-mass, high-metallicity host galaxies presented by \citet{Levesque10} and other studies \citep{Fynbo06, Prochaska07, Fynbo08, Chen08}, challenge the notion of a sharp metallicity threshold \citep{Modjaz06, Kocevski09} for the host galaxies that are capable of producing LGRBs.  Moreover, without such a strict low metallicity criteria for the generation of GRB progenitors, the reason of the preference of GRBs to occur in environments with relatively lower chemical enrichment remains unclear.

In this letter we examine the mass-metallicity relation for GRB host galaxies by investigating the effects of a recently proposed anti-correlation between a galaxy's star formation rate and its metallicity. By examining star forming galaxies from a number of catalogs, \citet{Mannucci10} found a trend of increasing SFR as a function of decreasing metallicity in low-mass galaxies. While this trend is prominent in low-mass galaxies, the significance of the anti-correlation decreases sharply with increasing galaxy mass, almost disappearing completely for high-mass galaxies. 

We show that such a connection between the SFR and chemical enrichment of a galaxy would naturally explain the systematic offsets reported by \citet{Han10} and \citet{Levesque10}.  If low metallicity galaxies produce more stars than their equally massive, high-metallicity counterparts, then it would be expected that transient events that closely trace the star formation history of their host galaxies would be more likely to occur in these high SFR, low-metallicity galaxies.  We test this hypothesis by modeling the $M$-$Z$ relation for LGRBs using a combination of data from the SDSS-DR7 catalog and model prescription for the SFR-$Z$ relation in Section 2.  We compare our model predictions to published host mass data in Section 3 and we discuss the implications of our results in Section 4.


 
\section{Model Prescriptions} \label{sec:Models}

\begin{figure}
\includegraphics[scale=0.8,width=\columnwidth, angle=0]{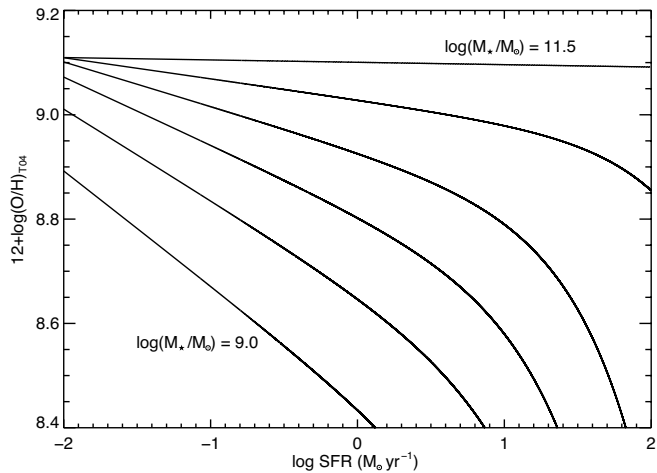}
\caption{An attenuated power-law model for the SFR-$Z$ relation for values of log($M_{\star}/M_{\odot})$ between $9.0 \leqslant$ log($M_{\star}/M_{\odot}) \leqslant 11.5$ at intervals of 0.5 log($M_{\star}/M_{\odot}$) .  The median SFR increases sharply as a function of falling metallicity for low mass galaxies, but no such relation is observed for their high mass counterparts.   
\label{Fig:SFRvsZ}}
\end{figure}

In order to investigate the nature of the metallically offset in the $M$-$Z$ relation for LGRB host galaxies, we must consider the difference between the probability distributions for the metallicity of a galaxy drawn from the general star forming population and a galaxy that is likely to harbor a LGRB.  For the general star forming population, the metallicity probability distribution $P(Z)$ for a galaxy of a given stellar mass $M_{\star}$ is simply the normalized metallicity distribution $\phi(Z,M_{\star})$ of all star forming galaxies at that mass.  Therefore, the most likely metallicity of a galaxy drawn randomly from the star forming population will reflect the peak of $\phi(Z,M_{\star})$. The most likely metallicity of a galaxy selected because of the occurrence of a LGRB is quite different.  For host galaxies, any metallicity dependance on the probability of the transient event, $P_{\rm GRB}(Z,M_{\star})$, must be taken into consideration.  In such a case, the resulting LGRB metallicity distribution would reflect the product of the metallicity distribution for all star forming galaxies and this metallicity dependent probability distribution, $P_{\rm Host}(Z) \sim \phi(Z,M_{\star}) * P_{\rm LGRB}(Z,M_{\star})$. 

The metallicity dependance on the probability of the transient event can represent a number of different theoretical considerations.  It could reflect, for example, the metallicity dependance on the likelihood that a star would possess the physical conditions required to produce a GRB.  Conversely, it could even reflect an observational bias such as the hypothetical probability of optically detecting the event in galaxies with metallicity dependent degrees of dust obscuration. In this letter we focus on the metallicity dependance of a galaxy's star formation rate as the primary contributor to $P_{\rm LGRB}(Z,M_{\star})$.

Because the production (and death) of massive stars is closely linked to the ongoing star formation in a galaxy, we assume that a higher SFR in a galaxy will result in a greater likelihood for the occurrence of a LGRB.  Therefore we set $P_{\rm LGRB}(Z,M_{\star}) \sim {\rm SFR}(Z,M_{\star})$.  Here we are explicitly assuming that an equal fraction of stars being produced, in both low and high SFR galaxies, contribute to the production of massive LGRB progenitors.  

\begin{figure}
\includegraphics[width=\columnwidth, angle=0]{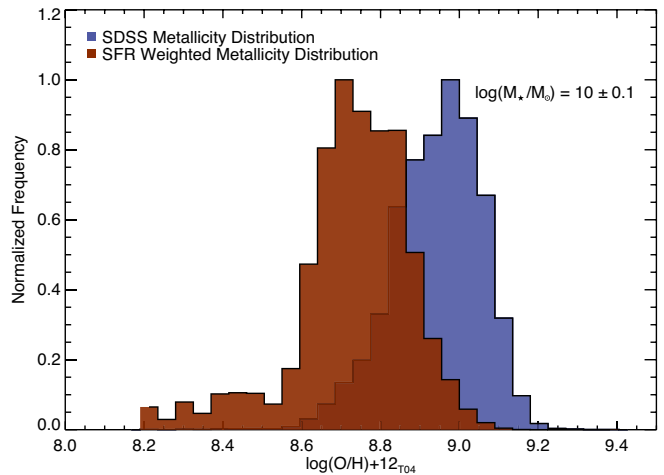}
\caption{The normalized metallicity distribution $\phi(Z)$ (blue histogram) for the general star forming population and the SFR weighted metallicity distribution $P_{\rm Host}(Z,M_{\star})$ distribution (red histogram) between $9.9 <$ log($M_{\star}/M_{\odot}) < 10.1$.  The peak of the $P_{\rm Host}(Z)$ distribution is distinctly offset to lower metallicities because of the greater weight given to low metallicity galaxies due to their higher SFR compared to their higher metallicity counterparts.  
\label{Fig:ConvolvedZSFR}}
\end{figure}

\begin{figure*}
\begin{center}
    \includegraphics[scale=0.8]{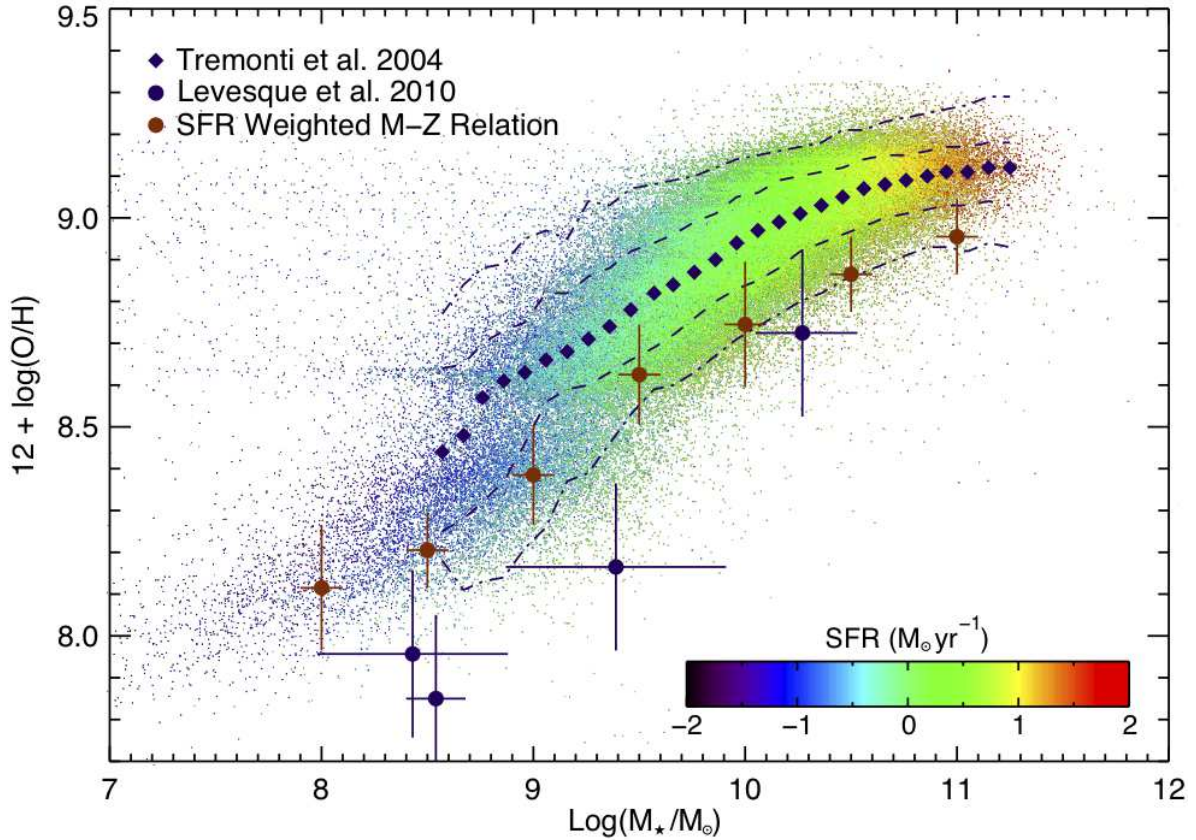}
   \caption{The M$-$Z relation for the general star forming population between $0.005 < z < 0.30$, with the color coding representing the absolute SFR.  The blue diamonds represent the median M$-$Z relation presented in \citet{Tremonti04} with the dashed and dotted lines representing the contours which include 68$\%$ and 95$\%$ of the data, respectively.  The red circles represent our predicted median GRB host galaxy metallicity as a function of log($M_{\star})$, effectively producing an SFR weighted $M$-$Z$ relation which is shifted to lower metallicities.  The vertical error bars represent the 1$\sigma$ spread in the underlying $P_{\rm Host}(Z)$ distribution.  The blue circles represent the mass and metallicity of GRB host galaxies selected from \citet{Levesque10} for which a conversion between the KK04 and T04 metallicity diagnostics was possible.}
   \label{Fig:HostMZ}
\end{center}   
\end{figure*}

In order to model the $M$-$Z$ relation for GRB host galaxies, we must convolve the $\phi(Z)$ and $P_{\rm LGRB}(Z)$ probability distributions and find the peak of the resulting $P_{\rm Host}(Z)$ distribution for a range of galaxy masses.  This requires adopting prescriptions for both $\phi(Z)$ and SFR($Z$) as a function of galaxy mass.  We can measured $\phi(Z)$ directly through the use of data from the MPA/JHU catalog\footnote{http://www.mpa-garching.mpg.de/SDSS} from SDSS-DR7\footnote{http://www.sdss.org/dr7}.  This catalog includes stellar masses and emission line estimates for over 927552 galaxies, providing a wealth of information regarding galaxy demographics in the local Universe.  The metallicity values presented in the catalog reflect oxygen abundance estimates based on a statistical method discussed in \citet{Tremonti04}.  We also utilize the catalog's stellar masses, obtained through broad-band fits SDSS photometry, and SFR estimates based on the technique discussed in \citet{Brinchmann04}.

For the purposes of our model, we selected emission line galaxies with a redshift cut of $0.005 < z < 0.30$ for which stellar mass estimates were possible.  For the remaining galaxies, we set a threshold to the signal-to-noise ratio (S/N) of H$\alpha$ of S/N $> 5$ in order to insure reliable metallicity estimates.  Metallicities for AGNs are not reported in the catalog and hence are not included in our sample.  Our final sample contains 155564 galaxies.  


For an analytic description of  SFR($Z$), we turn to \citet{Mannucci10}, where the authors quantified a trend between SFR and metallicity using the SDSS-DR7 MPA/JHU catalog at redshifts between 0.07 and 0.30.  They find that for all log($M_{\star}) <10.7$, SFR increases with decreasing metallicity at constant mass (e.g. see their Figure 1,  \citet{Mannucci10}).  They provide a 4th order polynomial fit to the observed SFR-$Z$ relation, although for the purposes of our analysis we have reworked their expression into an attenuated power-law in order to extrapolate their original fit below log(SFR) $< -1$ and above log(SFR) $> 1$.  Our analytic model for SFR($Z$) is presented in Figure \ref{Fig:SFRvsZ} for several different values of log($M_{\star}/M_{\odot})$ between $9.25 \leqslant$ log($M_{\star}/M_{\odot}) \leqslant 11.35$.  We adopt this empirical model as the prescription for SFR($Z$) as a function of $M_{\star}/M_{\odot}$.

\begin{figure*}
\begin{center}
    \includegraphics[scale=0.80]{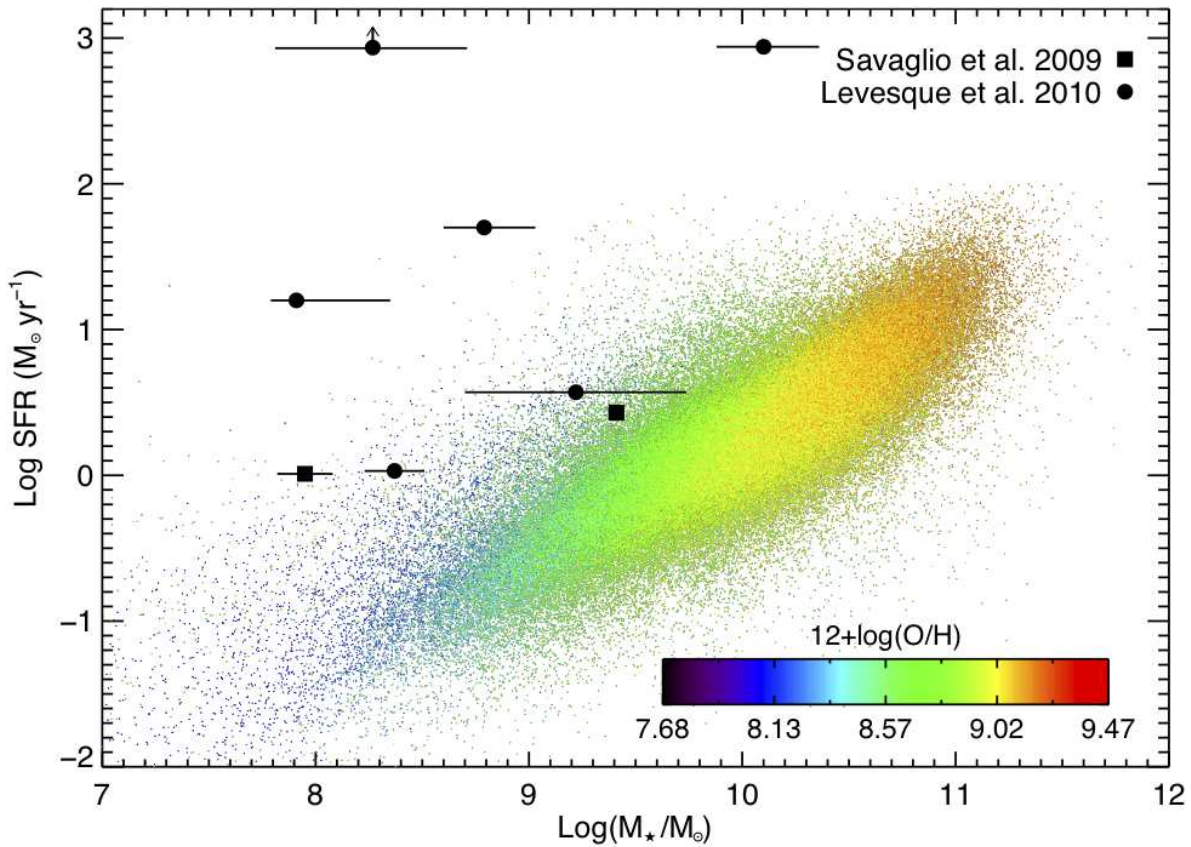}
   \caption{The SFR vs. $M_{\star}/M_{\odot}$ for all 155564 galaxies in our SDSS sample along with 6 galaxies presented in \citet{Levesque10} and 2 galaxies from \citep{Savaglio09} that fell below $z < 0.3$.  The color gradient represents metallicity and the expected trend of increasing SFR and metallicity as a function of galaxy mass is clearly evident.  All 8 of the low redshift LGRB hosts under consideration lie to the upper left of the SFR-$M_{\star}/M_{\odot}$ trend.}
   \label{Fig:SFRComparison}
\end{center}   
\end{figure*}

\section{Results} \label{sec:Results}

Using the subset of data from MPA/JHU catalog, we can obtain $\phi(Z)$ for effectively a constant mass by selecting all galaxies within a narrow $M_{\star}/M_{\odot}$ range.  For example, in Figure \ref{Fig:ConvolvedZSFR} we show the normalized $\phi(Z)$ (blue histogram) selected for all galaxies between $9.9 <$ log($M_{\star}/M_{\odot}) < 10.1$ $\phi(Z)$, totaling 20450 galaxies.  The peak of this distribution represents the median metallically for a star forming galaxy in the MPA/JHU catalog for log($M_{\star}/M_{\odot}$) $\sim 10.0 \pm 0.1$.  We can then calculate the median metallicity of a galaxy likely to host a GRB by weighting $\phi(Z,M_{\star})$ by SFR($Z$,$M_{\star}$) and finding the peak of the resulting probability distribution $P_{\rm Host}(Z,M_{\star})$.  The normalized $P_{\rm Host}(Z,M_{\star})$ distribution is shown in Figure \ref{Fig:ConvolvedZSFR} (red histogram).  The peak of the $P_{\rm Host}(Z,M_{\star})$ distribution has shifted to lower metallicities because of the greater weight given to low metallicity galaxies due to their higher SFR compared to their higher metallicity counterparts.

We calculate the median of the $\phi(Z)$ and $P_{\rm Host}(Z)$ distributions for a range of log($M_{\star}/M_{\odot}$) values between $8.0 <$ log($M_{\star}/M_{\odot}) < 11.0$ and plot the results in Figure \ref{Fig:HostMZ}.  The red circles represent the predicted median GRB host galaxy metallicity as a function of log($M_{\star}/M_{\odot})$, effectively producing an SFR weighted $M$-$Z$ relation that is shifted to lower metallicities compared to the $M$-$Z$ relation for the general star forming galaxy population.  The vertical error bars represent the 1$\sigma$ spread in the underlying $P_{\rm Host}(Z)$ distribution.  The offset between the SFR weighted and general $M$-$Z$ relations is roughly $\sim0.3$ at log($M_{\star}/M_{\odot}$) $\sim 9.0$ and $\sim0.15$ at log($M_{\star}/M_{\odot}$) $\sim 11.0$.  The decrease in the offset between the SFR weighted and general $M$-$Z$ relations is consistent with the trend reported by \citet{Levesque10} in which the metallicities of high-mass host galaxies are more consistent with the general $M$-$Z$ relation compared to their low-mass counterparts.  

We compare our modeled SFR weighted $M$-$Z$ relation to the mass and metallicity of a subset of GRB host galaxies selected from \citet{Levesque10}.  Because \citet{Levesque10} report their metallicity values in the KK04 system, we convert their reported value to the T04 system using the conversion tables provided in \citet{KK04}.  Since this conversion is not defined for all metallicities, we selected a subsample of 4 host galaxies below $z < 0.3$ that fell within the metallicity range for which this conversion was possible.  

Although the predicted SFR weighted $M$-$Z$ relation is more consistent with the data than to the general $M$-$Z$ when considering the errors in both mass and metallically, all four galaxies still nonetheless systematically fall below our model. We return to the implications of this discrepancy in the next section.

\section{Discussion} \label{sec:Discussion}

The nature of the low metallicity offset in the GRB defined $M$-$Z$ relation reported by \citet{Levesque10} is rather intriguing.  The observed shift supports previous spectroscopic evidence that GRBs tend to preferentially occur in low metallicity environments, although the relative nature of the offset, in addition to several high metallicity host galaxies reported by \citet{Levesque10} and \citep{Savaglio09}, suggests that host galaxies do not necessarily adhere to a strict metallicity ``cut-offÓ. 


The possibility of a correlation between the chemical enrichment and the rate of star formation in a galaxy may begin to explain the nature of this relative offset, without the need of such a cut-off in terms of absolute metallicity.  If low metallicity galaxies tend to produce more stars than their high metallicity counterparts for a given stellar mass, then the likelihood of detecting a transient event linked to a massive, short lived progenitor from within a low metallicity galaxy will increase accordingly.  Therefore the nature of the offset of the GRB defined $M$-$Z$ relation towards lower metallicity could largely be explained as a bias towards finding transient events in galaxies with relatively higher SFRs.  

Several studies have already indicated that LGRBs do tend to occur in galaxies with higher specific star formation than typical star forming galaxies at similar redshifts \citep{Chary02, Berger03, Christensen04, Savaglio09}.  We illustrated this in Figure \ref{Fig:SFRComparison}, where we plot the SFR vs. $M_{\star}/M_{\odot}$ for all 155564 galaxies in our SDSS sample along with 6 galaxies presented in \citet{Levesque10} and 2 galaxies from \citep{Savaglio09} that fell below $z < 0.3$.  The color gradient represents the metallicity of the SDSS galaxies and the expected trend of increasing SFR and metallicity as a function of galaxy mass is clearly evident.  All 8 of the low redshift LGRB hosts under consideration occupy the upper left of the SFR-$Z$-$M_{\star}/M_{\odot}$ plot, having SFRs that are substantially larger than the mean SDSS sample for their galaxy mass.  These few galaxies illustrate the degree to which these host galaxies have SFRs that exceed the star formation activity of similar galaxies.

A bias towards finding transient events in high SFR galaxies alone does not explain why LGRBs occur in small to intermediate size galaxies as opposed to the most massive galaxies, which typically have much higher absolute rates of star formation.  To understand the mass distribution of LGRB host galaxies, we must consider the number of galaxies as a function of mass (the galaxy mass function).  Although high mass galaxies produce far more stars than their low mass counterparts at all redshifts, the number of low mass galaxies in the Universe exceeds that of high mass galaxies by several orders of magnitude.  As a result, the SFR weighted galaxy mass function, i.e. the galaxy mass at which most of the star formation is occurring at a given redshift, peaks at intermediate masses.  Therefore the fact that we do not see LGRBs occurring predominately in the most massive, highly star forming, galaxies is not in contradiction to the suggestion that transient selected galaxies are biased towards high SFR environments.

In \citet{Kocevski09} we compared the mass distribution of GRB host galaxies to the SFR weighted galaxy mass function.  Using a similar argument as we have in this letter, we modeled the SFR weighted galaxy mass function for a variety of redshifts by convolving prescriptions for the the galaxy mass function $\phi(M_{\star}/M_{\odot})$ at a given redshift with the SFR as a function of stellar mass.  We found that the SFR weighted mass function at $z = 1$ peaks at roughly log($M_{\star}/M_{\odot}$) $\sim$ 10.3.  The measured median stellar mass of LGRB host galaxies reported by \citet{Savaglio09} for the same redshift is slightly lower, roughly log($M_{\star}/M_{\odot}$) $\sim$ 9.3.  


We interpreted this discrepancy as evidence for a metallicity cut-off at which the SFR weighted galaxy mass function would be truncated due to the relationship between a galaxy's mass and metallicity, resulting in a host mass distribution which would be shifted to lower masses.  Such a metallicity cut-off may not be necessary if in fact low mass, low metallicity galaxies produce more stars than their high metallicity counterparts.  In such a scenario, the SFR weighted galaxy mass function would be shifted to lower masses without the need of a metallicity cut-off due to the additional weight given to low mass, low metallicity galaxies.  The effects of a SFR-$Z$ relation for low mass galaxies was not taken into account in the analysis performed by \citet{Kocevski09} and a full examination of the effects of the correlation on the resulting joint mass and metallicity distributions will be reserved for a future paper. 

Although we have shown that a metallicity dependance on the SFR of low mass galaxies can in fact produce an $M$-$Z$ relation offset to lower metallicities, the mass and metallicity data taken from \citet{Levesque10} nonetheless systematically fall below our modeled $M$-$Z$ relation.  Within the context of our model, this discrepancy may be evidence for a further bias in the probability of LGRBs to occur in relatively low metallicity environments beyond the metallicity dependent SFR that we have considered.  A sharp decrease in $P_{\rm LGRB}(Z)$ with increasing metallicity due to a metallicity dependance on the physical conditions required to produce a GRB, would be one such effect.  In the framework of the collapsar model, low metallicity progenitors would retain more of their mass due to the reduction of line-driven stellar winds \citep{Kudritzki00, Vink05}, and hence preserve more of their angular momentum and stellar mass at the time of collapse \citep{Yoon05, WoosleyHeger06}.  This mechanism may be crucial to the production of collimated emission that is thought to be required to produce LGRBs. 

Therefore, although the observations of several high mass, high metallicity LGRB host galaxies may rule out a strict metallicity ``cut-off", the discrepancy between the host galaxy data measured by \citet{Levesque10} and our modeled $M$-$Z$ relation may point to further metallicity biases in the LGRB sample.  These biases need only to reflect a decreasing likelihood of a GRB occurrence as a function of increasing metallicity in order to provide a mechanism to further shift the LGRB defined $M$-$Z$ relation to lower metallicities.  Such biases would compound the effect introduced through the simple metallicity dependent SFR prescription that we have assumed in our model and as such there would be no need for a strict metallicity ``cut-off" to explain the relative preference for LGRBs to occur in low metallicity environments. 

Finally, our work predicts that core collapse SNe found through untargeted surveys such as the Palomar Transient Factory (PTF), Pan-Starrs, and the Large Synoptic Survey Telescope (LSST) should also define an $M$-$Z$ relationship that is shifted to lower metallicities compared to the general star-forming galaxy population.  Such an effect should not be expected of transient events that do not closely follow the star formation history of their host galaxies, such as short duration GRBs and SN Ia events.  If subsequent observations of SN II and Ibc host galaxies found through these surveys do not show such an offset, then the shift in the GRB defined $M$-$Z$ relation would necessarily be a physical effect intrinsic to GRBs and not due to the SFR-$Z$ selection bias that we propose. 

\section{Conclusions} \label{sec:Conclusions}

A growing body of spectroscopic evidence has made it clear that GRBs tend to preferentially occur in host galaxies that exhibit lower metallicities compared to the general star-forming galaxy population.  Recent stellar mass and metallicity estimates made by \citet{Levesque10} indicate that host galaxies follow a $M$-$Z$ relation that is offset to lower metallicities compared to the general star-forming galaxy population.  \citet{Levesque10} find that this offset, while pronounced for low mass galaxies, becomes negligible for two GRBs found to occur in high-mass, high-metallicity host galaxies.  

We show that the nature of this reported offset may be explained by a recently proposed anti-correlation between the SFR and the metallicity of star forming galaxies. If low metallicity galaxies produce more stars than their equally massive, high-metallicity counterparts, then it would be expected that transient events that closely trace the star formation in a galaxy would be more likely to be found in these low metallicity, low mass galaxies.  We quantify this hypothesis by modeling the $M$-$Z$ relation for GRBs.  We calculate the SFR weighted metallicity distribution for star forming galaxies by using metallicity data from the MPA/JHU catalog as part of SDSS-DR7, as well as a model prescription for the relationship between a galaxy's SFR and metallicity reported by \citet{Mannucci10}.  We find that the SFR weighted metallicity distribution peaks at significantly lower metallicity compared to the metallicity distribution for the general star forming galaxy population at a given stellar mass.  

Therefore, the offset between the GRB $M$-$Z$ relation and the properties of the general star forming population could simply be due to the methods used to select their respective galaxy populations, with GRBs being biased towards low metallicity, high SFR, galaxies.  Although our model provides a framework to explain the nature of this low metallicity offset, we find that further biases beyond a metallicity dependent star formation may be required to account for the full degree of the discrepancy between the GRB and SDSS defined $M$-$Z$ relation.

\acknowledgements

D.K. acknowledges financial support through the Fermi Guest Investigator program.  We thank Maryam Modjaz, Christi Tremonti, Emily Levesque, Daniel Perley, and Risa Wechsler for insightful discussions. This work was supported in part by the NASA Fermi Guest Investigator program under NASA grant number NNX10AD13G and the U.S. Department of Energy contract to the SLAC National Accelerator Laboratory under no. DE-AC3-76SF00515.


\end{document}